\documentclass[twocolumn]{aa}
\usepackage{graphicx}

\def\etal{{\it et al.}}
\def\llabel#1{\label{#1}} 

\def\vii{{\bf i}}
\def\vjj{{\bf j}}
\def\vkk{{\bf k}}
\def\vII{{\bf I}}
\def\vJJ{{\bf J}}
\def\vKK{{\bf K}}
\def\vKK{{\bf K}}

\def\r{{\bf r}}
\def\v{{\bf v}}

\def\cI{{\cal I}}
\def\cP{{\cal P}}
\def\cS{{\cal S}}
\def\cM{{\cal M}}

\def\om{\omega}
\def\O{\Omega}
\def\cR{{\cal R}}
\def\a{\alpha} 

\def\Frac#1#2{{{\displaystyle\strut#1}\over{\displaystyle\strut#2}}}

\def \be  {\begin{equation}}
\def \ee  {\end{equation}}

\def\Rotx#1{\left(\matrix{1& 0& 0\cr 0& \cos#1& -\sin#1\cr
                   0& \sin#1& \cos#1\cr}\right)}

\def\Rotz#1{\left(\matrix{\cos#1&-\sin#1&0\cr\sin#1&\cos#1&0\cr
                               0& 0& 1\cr}\right)}
\def\Vect#1#2#3{\left(\matrix{#1\cr#2\cr#3\cr}\right)}

\def\EQM#1{\vcenter{\normalbaselines{\mathsurround=0pt}
    \ialign{${\displaystyle ##}$\hfil&&\ ${\displaystyle ##}$\hfil\crcr
    \mathstrut\crcr\noalign{\kern-\baselineskip}
    \noalign{\smallskip}
    #1\crcr\mathstrut\crcr\noalign{\kern-\baselineskip}}}}

\def\figsiz{8cm}

\newcommand\figa{
\begin{figure}[h] 
\begin{center}
\begin{tabular}{c c}
 \includegraphics[width=\figsiz]{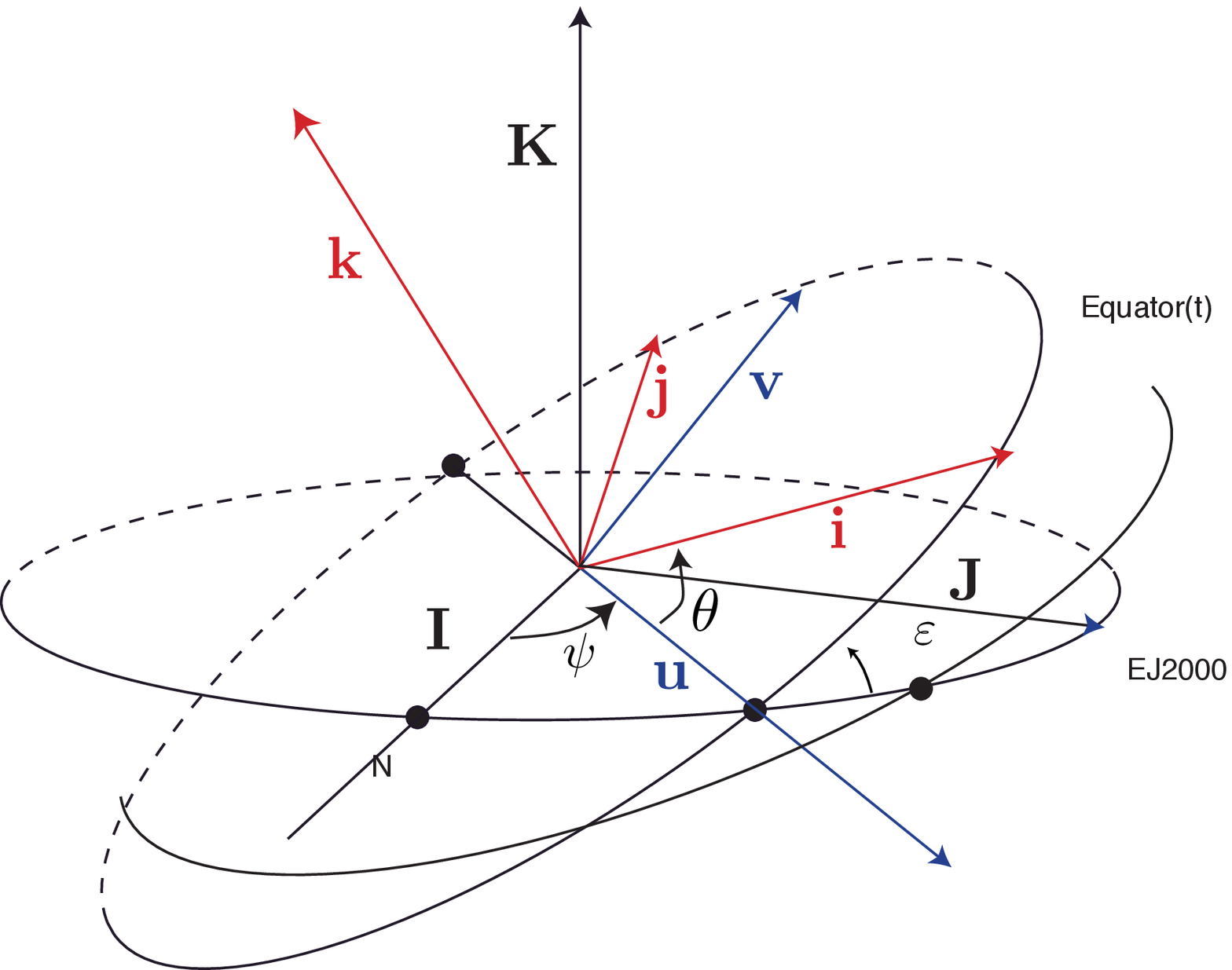} & 
 \end{tabular}
  \caption{ 
  } 
  \llabel{Figa}
\end{center}
\end{figure}
}

\title{A comment on `Accurate spin axes and solar system dynamics:
 Climatic variations for the Earth and Mars' }
\author{J. Laskar }
\authorrunning{J. Laskar }
\titlerunning{A comment on `Accurate spin axes and solar system dynamics' }

\institute{ Astronomie et Syst\`emes Dynamiques, IMCCE-CNRS UMR8028, 
77 Av. Denfert-Rochereau, 75014 Paris, France }
\offprints{J. Laskar, \email{laskar@imcce.fr}}

\date{\today}
\abstract{
In a recent paper, Edvardsson \etal (\cite{edvardsson}) propose a new solution for the spin 
evolution of the Earth and Mars. Their results  differ significantly with respect to previous studies,
as they found a large contribution on the precession of the planet axis 
from the tidal effects of Phobos and Deimos. In fact, this probably results from 
the omission by the authors of the torques exerted on the satellites orbits by the planet's 
equatorial bulge, as otherwise the average torque exerted by the satellites on the planet is null. }
  
\begin{document}
\maketitle

\section{Introduction}

In a recent paper, Edvardsson \etal (\cite{edvardsson}) propose a new solution for the spin 
evolution of the 
Earth and Mars. 
Considering the absence of a precise evaluation of the errors due to the integrator 
and the absence of relativity in the model,
one could discuss the use of the word 'accurate' in the  title of the paper,
but unfortunately  there are some  more important 
flaws in this paper. In their integration of the spin of Mars, the authors found
 that the integration of the spin changes 
in a large amount when Phobos and Deimos are taken into account (see their Fig.12). They also notice that 
their " curve without the moons is very similar to the curve given 
by Bouquillon \& Souchay (\cite{bouquillon}) ", who included the moons. 
In fact,  the  paper of Edvardsson \etal (\cite{edvardsson}) is largely in error on this point.

\section{Precession due to a distant satellite}
\figa

Let us consider a planet $\cP$   with momentum of inertia $ A = B < C$ orbiting the Sun on 
a fixed ellipse (we will not consider here the planetary perturbations or the perturbations due to the 
satellite presence), and a satellite  $\cS$  of mass $m$ orbiting the planet. Let $(\vii,\vjj,\vkk)$   be 
a basis linked to $\cP$ with $\vkk$ associated to the axis of maximum inertia $C$. Let $(\vII,\vJJ,\vKK)$
be a fixed reference frame, with origine in the direction of $\vII$, and $\vKK$ normal to the 
orbital plane of $\cP$ (Fig.1). $a$ and $e$ are the semi-major axis and eccentricity of $\cS$, $\Omega$ and $i$
the longitude of the axcending node  of the satellite orbits over the orbital plane of the planet $(\vII,\vJJ)$,
while $v$ is the true longitude, and $\omega$ the argument of perihelion. If $\r$ is the radius vector from the 
planet's to the satellite's center of mass,  with modulus $r$ and unitary vector $\v = \r /r$, the torque 
exerteed by $\cS$ on $\cP$ is 
\be
\Gamma = \Frac{3 G m}{r^3} \v \wedge \cI \v
\ee
where $G$ is the gravitational constant and $\cI$ the matrix of inertia ($\cI = {\rm diag} (A,A,C)$). Noting that 
$\cI = {\rm diag} (A,A,A)- {\rm diag} (0,0,C-A)$, (1)  can also be expressed as (see Murray, 1983)
\be
\Gamma = -\Frac{3 G m}{r^3} (C-A)\vkk \wedge (\v \v^\tau) \vkk
\ee
where $\v^\tau$ denotes the transposed of $\v$ ($\v^\tau  \vkk$ 
is thus the dot product of the two vectors $\v$ and $\vkk$). We will 
average $\Gamma$ over the fastest angle of this problem, that is over the mean anomaly $M$ of the satellite 
(the rotational angle $\theta$ is removed  after the asumption $A=B$). As $\vkk$ does not depends on $M$, 
the only expression to average in (2) is 
$
\gamma =  {\v \v^\tau}/{r^3}$.
In the $(\vII, \vJJ, \vKK)$ frame, the coordinates of $\v$ are 
\be
\v = \cR_3(\O) \cR_1(i)  \Vect{\cos(v+\om)}{\sin(v+\om)}{0}
\ee
with 
\be
\matrix{
\cR_1(\a) = \Rotx{\a} \ ;\cr
\noalign{\medskip}  
\cR_3(\a) = \Rotz{\a} \ .}
\ee

As $r^2\, dv = a^2\sqrt{1-e^2} dM$, after averaging over $M$, we obtain
\be
<\gamma>_M = \Frac{1}{2a^3(1-e^2)^{3/2}} [ Id - \cM]
\ee
where $\cM$ is expressed in $(\vII,\vJJ,\vKK)$ as the $3\times 3$ matrix 
\be
\cM  = \cR_3(\O) \cR_1(i) \left(\matrix{0& 0& 0\cr 0& 0& 0\cr
                   0& 0& 1\cr}\right)\cR_1(-i)\cR_3(-\O) 
\ee

When the precession of the node is rapid with respect to the precession of the 
spin axis of the planet (as for the Moon around the Earth), one can also average over $\Omega$. 
We have thus in the basis $(\vII,\vJJ,\vKK)$ 
\be
<\cM>_\O = \frac{\sin^2i}{2} Id  + (1-\Frac{3\sin^2i}{2})\left(\matrix{0& 0& 0\cr 0& 0& 0\cr 0& 0& 1\cr}\right)
\ee

Finally, as the parts involving the identity $Id$ cancel, we are left with 
\be
\EQM{
<\Gamma>_{M,\O} &= -\Frac{3 G m(C-A)}{2a^3(1-e^2)^{3/2}}  (1-\Frac{3\sin^2i}{2})\vkk \wedge \vKK \vKK^\tau \vkk \cr
&= -\Frac{3 G m(C-A)}{2a^3(1-e^2)^{3/2}}  (1-\Frac{3\sin^2i}{2})\cos \varepsilon\ \vkk \wedge \vKK }
\ee
which leads to the classical contribution  $p_s$ of the precession of spin axis  $\vkk$ around the normal
to the orbital plane
$\vKK$, when the rotational rate $\nu$ is large with respect to the precession frequency $p$.
\be
p_s = -\Frac{3 G m}{2\nu \,a^3(1-e^2)^{3/2}}\Frac{C-A}{C}  (1-\Frac{3\sin^2i}{2})\cos \varepsilon
\ee

\section{Precession due to a close satellite}
The previous study is in fact valid uniquely in the case of a  satellite sufficiently far from the planet, 
so that the precession of its orbit is mostly driven by the solar perturbations, and not by the 
torque exerted by the equatorial bulge of the planet. In the case of a close satellite, it was 
shown  (Goldreich, \cite{goldreich}, Kinoshita, \cite{kinoshita}) that instead of precessing  around 
the normal $\vKK$ to the orbital plane $(\vII, \vJJ)$,
with a roughly constant inclination $i$ with respect to  $(\vII, \vJJ)$, the satellite will precess 
with respect to $\vkk$ with a nearly constant inclination $i' $ with respect to the equator plane $(\vii, \vjj)$.
In this case, all the previous study is still valid, but the orbit of the satellite is now referred to 
the equatorial reference frame $(\vii,\vjj,\vkk)$, and the second averaging is made with respect to the 
longitude of the  node,  $\O'$, of the satellite orbit with respect to the equatorial plane $(\vii,\vjj)$.
The average expression (6, 7) is still valid, when replacing $\O$ and $i$ by $\O'$ and $i'$. The 
average $<\cM>_{\O'}$ will then be given by (8), after changing $i$ with $i'$, but this expression is now 
obtained in the reference frame of the planet $(\vii,\vjj,\vkk)$. As $\vkk \wedge \vkk \vkk^\tau \vkk =\vkk \wedge \vkk =0 $,
we have now for the averaged torque 
\be
<\Gamma>_{M,\O'} =0 .
\ee

\section{Discussion}
Thus, contrarily to what is found by Edvardsson \etal (2002), the averaged torque exerted by Phobos and Deimos 
on Mars should be null (at first order). The origin of the authors' error can be traced by looking to their 
figures 1 and 2. The inclination of the satellites with respect to the equator start with a value close to zero at the origin, 
and then increase up to about 50 degrees and then back to zero with a period of about 6000 and 1500 years  for Phobos and  Deimos,
while a back on the enveloppe calculation gives
respectively 5400 and 1400 years for the precession of the node of these satellites submitted uniquely to solar perturbation. The issue is then clear. By neglecting the torque exerted on the satellites 
orbits by the equatorial bulge of Mars, the authors found that instead of remaining close to the equatorial plane of Mars,
as demonstrated by Goldreich (1965), the 
satellites were precessing along the orbital plane. Instead of finding a zero value for the mean precession torque (11), they 
obtain an effect following Eq. (10) which gives a contribution of about $0.5$"/year to the precession of mars axis,
in agreement with the difference observed by the authors in figure 12. It is surprising that the authors did 
not wonder how Phobos and Deimos could both be at present with an inclination of less than 2 degrees with respect to the equator 
if they were precessing along the orbit.  Goldreich indeed realised more that this `would amount to an unbelievable coincidence'.


\begin{thebibliography}{}
\bibitem[2002]{edvardsson} Edvardsson, S., Karlsson, K.G., Engholm, M. 2002, A\&A, 384, 689--701
\bibitem[1983]{murray} Murray, C. A. 1983,  Vectorial Astrometry, Adam Hilger Ltd, Bristol
\bibitem[1965]{goldreich} Goldreich, P. 1965, AJ, 70, 5--9
\bibitem[1993]{kinoshita} Kinoshita, H. 1993, Celest. Mech. 57, 359--368
\bibitem[1999]{bouquillon} Bouquillon, S. \& Souchay, J. 1999, A\&a, 345, 282
\end{thebibliography}
\end{document}